\documentclass[sigconf]{acmart}
\usepackage{enumitem}
\AtBeginDocument{%
  }

\copyrightyear{2026}
\acmYear{2026}
\setcopyright{cc}
\setcctype{by-nc-nd}
\acmConference[CIKM '26]{Proceedings of the 35th ACM International Conference on Information and Knowledge Management}{November 07--11, 2026}{Rome, Italy}
\acmBooktitle{Proceedings of the 35th ACM International Conference on Information and Knowledge Management (CIKM '26), November 07--11, 2026, Rome, Italy}
\acmDOI{10.1145/3799682.3841099}
\acmISBN{979-8-4007-2539-5/2026/11}

\usepackage{graphicx}
\usepackage{multirow}
\usepackage{booktabs}
\begin{document}

\title[LSF-SR: Latent Semantic Fusion for Sequential Recommendation via Flow-based Conditional Variational Autoencoders]{LSF-SR: Latent Semantic Fusion for Sequential Recommendation via Flow-based Conditional Variational Autoencoders}


\author{Shih-Hong Chen}
\email{shihhong25.cs13@nycu.edu.tw}
\affiliation{%
  \institution{National Yang Ming Chiao Tung University}
  \city{Hsinchu}
  \country{Taiwan}
}

\author{Josh Jia-Ching Ying}
\email{jashying@gmail.com}
\affiliation{%
  \institution{National Chung Hsing University}
  \city{Taichung}
  \country{Taiwan}
}

\author{Vincent S. Tseng}
\authornote{Corresponding author.}
\email{vtseng@cs.nycu.edu.tw}
\affiliation{%
  \institution{National Yang Ming Chiao Tung University}
  \city{Hsinchu}
  \country{Taiwan}
}

\renewcommand{\shortauthors}{Shih-Hong Chen, Josh Jia-Ching Ying, and Vincent S. Tseng}

\begin{abstract}
Sequential recommendation aims to predict users' future interests from their historical interactions.
Although Large Language Models (LLMs) capture rich item semantics, existing methods often struggle to align collaborative signals with textual semantic knowledge.
As a result, the learned item representations fail to capture the complementary strengths of both signals, leading to suboptimal recommendation quality.
To address this limitation, we propose \textbf{L}atent \textbf{S}emantic \textbf{F}usion for \textbf{S}equential \textbf{R}ecommendation via Flow-based Conditional Variational Autoencoders (\textbf{LSF-SR}), a novel framework that uses a Conditional Variational Autoencoder (CVAE) with Normalizing Flows to fuse item ID embeddings and LLM-generated semantic signals.
At the core of LSF-SR is a conditional fusion module augmented with planar or radial flows.
This module learns a flexible latent space that encourages items with similar semantic profiles to cluster together within the latent manifold.
Through extensive experiments on five public benchmark datasets, we demonstrate that LSF-SR consistently outperforms state-of-the-art baselines, achieving gains of up to 12.98\% and 14.13\% in Recall@20 and NDCG@20, respectively.
\end{abstract}



\maketitle
\section{Introduction}
\label{sec:intro}
Sequential Recommendation (SR) has become a crucial aspect of modern information systems, as it aims to capture users' evolving preferences from their historical interactions \cite{sr_survey}. 
The main challenge is modeling the complex temporal dependencies in user behavior sequences to accurately predict what a user is likely to engage with next. 
In recent years, attention-based architectures have become prevalent in this field, particularly self-attention mechanisms, which effectively assign contextual weights to historical items and model long-range dependencies within sequences \cite{sasrec, bert4rec}. 
These models have set competitive performance benchmarks across various recommendation scenarios and remain the standard framework for sequential recommendation research.

Despite their effectiveness, attention-based sequential models face a significant limitation: they rely solely on discrete item identifiers for representation. 
In this framework, each item is mapped to a learnable embedding vector, and the semantic meaning of these vectors derives entirely from co-occurrence patterns in the training data. 
This approach works well when interaction data is abundant, making collaborative signals strong. 
However, it becomes fragile under two common conditions. 
First, when the overall interaction matrix is sparse, item embeddings train poorly, making it hard to capture reliable preference signals. 
Second, for new or rarely interacted items, the cold-start problem arises, making collaborative embeddings almost uninformative because they lack sufficient behavioral signals to inform their representations. 
These limitations pose significant challenges for real-world recommendation systems, where item catalogs are extensive, and interaction distributions often follow a long-tail pattern.

\begin{figure*}[t]
    \centering
    \includegraphics[width=\textwidth]{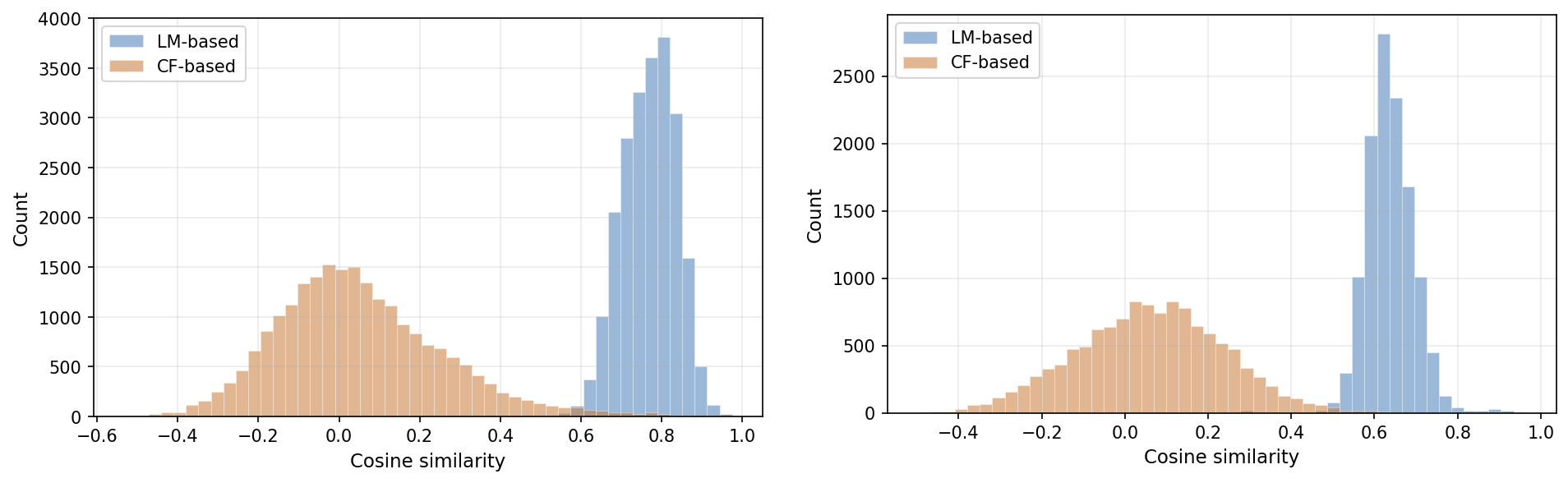}
    \Description{Left: cosine similarity distribution for a randomly selected user. Right: cosine similarity distribution for a randomly selected item. The distributions indicate strong misalignment between collaborative and LLM semantic embeddings.}
    \caption{Empirical analysis of the distributional mismatch between collaborative and LLM semantic embeddings on the Amazon Beauty dataset. We randomly select a user (left) and an item (right) and compute their cosine similarity distributions against all remaining users and items, respectively, revealing severe misalignment in their representations.}
    \label{fig:distribution_mismatch}
  \end{figure*}

To address the inherent constraints of sequential modeling, researchers have developed several approaches. 
A major focus is integrating structural side information—such as item categories, brands, or user ratings—into sequential models to improve performance on sparse interaction data \cite{diff, difsr, asif}. 
By incorporating these attributes, these side-information-integrated methods can strengthen item representations even when behavioral signals are limited.
However, these techniques typically rely on deterministic feature fusion methods, such as concatenation or gating mechanisms, which assume that side attributes are available, consistent, and complete. 
In practice, item metadata is often fragmented or missing, which can hinder the effectiveness of these approaches when attribute coverage is incomplete.
Another line of research addresses noise in behavioral sequences through frequency-domain filtering and refinements to attention mechanisms \cite{fmlprec, fearec, bsarec}.
\enlargethispage*{1\baselineskip}
These spectral methods convert interaction sequences to the frequency domain and apply selective filtering to reduce short-term behavioral fluctuations, revealing more stable underlying preference signals. 
Although these methods improve signal quality, they do not enrich the semantic content of item representations and thus fail to address the core issue of sparse collaborative signals.

In contrast, contrastive learning has emerged as an influential paradigm that uses self-supervised objectives to improve item representations. 
Conventional methods create alternative views of interaction sequences through data augmentation, such as masking, reordering, or cropping \cite{coserec, cl4srec}. 
More advanced techniques move beyond random perturbations by constructing high-quality contrastive pairs through coarse-to-fine intent clustering or hierarchical sequence organization \cite{iclrec, fenrec, icsrec, duorec, mclrec}. 
Although effective at enhancing representation robustness, contrastive learning operates solely within the collaborative embedding space. 
Regardless of how the contrastive pairs are formed or how sophisticated the augmentation strategies are, the resulting representations depend entirely on behavioral co-occurrence patterns. 
Therefore, although contrastive learning can improve the quality of collaborative signals, it cannot compensate for the lack of semantic knowledge about items. 
As a result, the semantic scope of these representations is limited to what can be inferred from interaction histories alone.

Recently, the impressive capabilities of Large Language Models (LLMs) in semantic understanding and contextual reasoning have inspired efforts to integrate rich semantic knowledge into sequential recommendation systems \cite{p5, unisrec}. 
By using LLMs to generate descriptive textual profiles from metadata, these methods add semantic grounding to item representations beyond mere item co-occurrence. 
Existing methods typically employ LLMs in one of two ways: first, as direct generative recommenders that reformulate the recommendation task as a text generation problem \cite{allmrec, alphafuse, llamarec, llm2rec}; or second, as tools for semantic extraction that generate item text embeddings, which are then used within collaborative filtering frameworks \cite{rlmrec, llmesr, lrd}. 
Recent techniques have also used LLM-generated content to define semantic neighborhoods that serve as self-supervised anchors for contrastive learning \cite{sracl}.

However, LLM-augmented recommendation methods face two fundamental and underexplored challenges. 
First, the semantic spaces generated by LLMs and the collaborative behavior spaces derived from interaction data are inherently incompatible. 
As shown in Figure~\ref{fig:distribution_mismatch}, LLM-generated embeddings tend to cluster tightly, exhibiting high inter-item cosine similarity. 
This clustering arises from the rich shared vocabulary and contextual overlap in natural language descriptions. 
In contrast, collaborative embeddings approximate a zero-centered Gaussian distribution, reflecting the stochastic and sparse nature of user behavior.
This distributional discrepancy creates disjoint neighborhood structures, so the top-$K$ neighbors identified in the textual semantic space often differ from those in the collaborative space. 
Consequently, attempts to combine these mismatched representations through late fusion or independent retrieval strategies often fail to align. 
Second, approaches that rely on LLM-based user-level profiling require continuous queries to the model during inference. 
Unlike item-level semantic extraction, which can be performed offline as a single preprocessing step, user-level profiling must be updated dynamically for each user at every inference step because user preferences evolve as their interaction histories grow. 
As a result, this approach introduces significant computational overhead that increases with the number of active users and the frequency of recommendation requests. 
Consequently, these methods may struggle to meet the real-time latency requirements needed for effective recommendations.

In response to these challenges, we propose \textbf{L}atent \textbf{S}emantic \textbf{F}usion for \textbf{S}equential \textbf{R}ecommendation via Flow-based Conditional Variational Autoencoders (\textbf{LSF-SR}), a novel probabilistic framework that addresses both distributional misalignment and the inference overhead of prior LLM-based approaches.
Rather than combining collaborative and textual semantic signals via deterministic projections or late fusion, LSF-SR models their interaction as a probabilistic inference problem.
We use a Conditional Variational Autoencoder (CVAE) \cite{sohn2015learning} in which LLM-derived semantic embeddings serve as conditioning signals.
These signals guide the transformation of collaborative item representations into a shared latent space.
This formulation accounts for uncertainty when aligning two heterogeneous representation spaces, enabling coherent joint representations even when the two modalities disagree.
To capture non-Gaussian structure in the joint distribution, we integrate normalizing flows \cite{rezende2015variational} into the approximate posterior, applying invertible transformations that reshape a simple base distribution into a more flexible form.
By restricting LLM invocations to the item level and running them offline, LSF-SR eliminates real-time computational overhead from user-level profiling while still benefiting from textual semantic signals.
Our main contributions are as follows:
\begin{itemize}[leftmargin=*, topsep=1pt, itemsep=1pt, parsep=0pt]
    \item We propose LSF-SR, a novel probabilistic fusion framework that directly addresses misalignment between collaborative behaviors and LLM-derived semantic signals.
    
    \item We develop a conditional fusion module with stacked normalizing flows to model a flexible posterior for precise semantic alignment.
    
    \item We implement a dynamic KL annealing strategy to prevent posterior collapse during training, ensuring that collaborative embeddings remain meaningfully grounded in semantic context.
    
    \item Our extensive experiments on five benchmark datasets \cite{yelp, amazon} show that LSF-SR consistently outperforms state-of-the-art methods, achieving up to 12.98\% in Recall@20 and 14.13\% in NDCG@20.
\end{itemize}

The remainder of this paper is organized as follows: We briefly review related work in Section 2 and present preliminaries in Section 3. In Section 4, we detail our proposed Latent Semantic Fusion for Sequential Recommendation via Flow-based Conditional Variational Autoencoders. Finally, Section 5 presents the results of our quantitative experimental evaluation, and Section 6 discusses our conclusions and future work.

\section{Related Work}
\label{sec:related}
To contextualize LSF-SR's contributions, we review the relevant literature across two interconnected research threads: conventional sequential recommendation, which underpins the behavioral modeling in our framework, and LLM-augmented recommendation, which motivates our semantic fusion strategy.

\subsection{Conventional Sequential Recommendation}
\label{sec:conventional_sr}
Conventional behavior-centric sequential recommendation models broadly fall into three research directions: attention mechanisms with frequency-domain enhancements, integration of side information, and contrastive learning.

\subsubsection{Attention Mechanisms and Frequency-Domain Enhancements.} 
Self-attention mechanisms are widely used in sequential recommendation because they weight past interactions by their relevance to predicting the next item \cite{sasrec, bert4rec}.
Some studies also incorporate frequency-domain denoising to reduce random behavioral noise.
These spectral techniques transform historical item sequences into the frequency domain, filter out short-term variations and misleading anomalies, and preserve the core signals of user preferences \cite{fmlprec, bsarec, fearec}.

\subsubsection{Side-Information Integrated Methods}
Integrating structural side information, such as item categories, brands, or user ratings, is a common strategy for alleviating data sparsity and cold-start issues in behavioral histories.
Early approaches primarily concatenated or gated these attributes with discrete behaviors at the input layer \cite{difsr, asif}. 
Over time, more advanced methods emerged that focus on operations in intermediate neural layers \cite{diff}. 
These strategies aim to preserve items' collaborative characteristics while leveraging structural attributes to enhance undertrained features in sparse interaction matrices.

\subsubsection{Contrastive Learning Paradigms.}
Contrastive learning has become increasingly popular as a robust approach to address the lack of supervisory signals through self-supervised representation learning.
Conventional contrastive sequential recommenders typically generate alternative views of interaction sequences using heuristic data augmentation \cite{cl4srec, coserec}.
More sophisticated methods adopt a coarse-to-fine approach to create high-quality contrastive pairs \cite{iclrec, duorec, icsrec, fenrec, mclrec}.
By organizing sequences or intent clusters hierarchically, these frameworks optimize multi-level contrastive objectives during training, effectively augmenting supervisory signals without compromising behavioral consistency.  

\subsection{LLM-Augmented Recommendation}
\label{sec:llm_augmented}
The remarkable semantic understanding and reasoning abilities of LLMs have sparked significant interest in improving behavior modeling.
In prevailing frameworks, LLMs are deployed in two distinct capacities: as direct generative recommenders or as text-based semantic extraction tools that support conventional models.
As direct recommenders, existing schemes reframe recommendations as a text-generation task, turning interaction histories into textual prompts to generate suggestions based on pre-trained knowledge or instruction tuning.
However, these methods are constrained by strict context-window limits and prohibitive real-time token inference costs, which inherently prevent them from capturing global collaborative behavior patterns.

Researchers are exploring a scalable approach that uses LLMs as advanced semantic extractors to improve conventional recommendation systems.
For instance, RLMRec \cite{rlmrec} combines collaborative representations with LLM-generated profile semantics by maximizing cross-view mutual information.
LRD \cite{lrd} leverages LLMs to extract language-based knowledge representations from textual metadata and uses a discrete-state variational autoencoder (DVAE) to autonomously discover latent item relations, which are jointly optimized with the sequential recommendation task.
Additionally, LLM-ESR \cite{llmesr} directly extracts frozen semantic embeddings from LLMs, employing a dual-view modeling approach and retrieval-augmented self-distillation to enhance representations for long-tail items and users in sparse scenarios.
More recently, SRA-CL \cite{sracl} uses LLMs to infer user preferences and item contexts, combining semantic retrieval with a learnable sample synthesizer and semantic item substitution to create high-quality positive pairs for contrastive learning within and across users.

Unlike prior strategies that rely on separate optimization steps or fixed, deterministic mappings, our LSF-SR framework introduces a fundamentally new paradigm. We frame the integration of behavioral and textual semantic signals as a flexible probabilistic inference task and use normalizing flows to reshape the posterior distribution, achieving coherent semantic alignment.

\begin{figure*}[t] 
    \centering
    \includegraphics[width=1.0\textwidth]{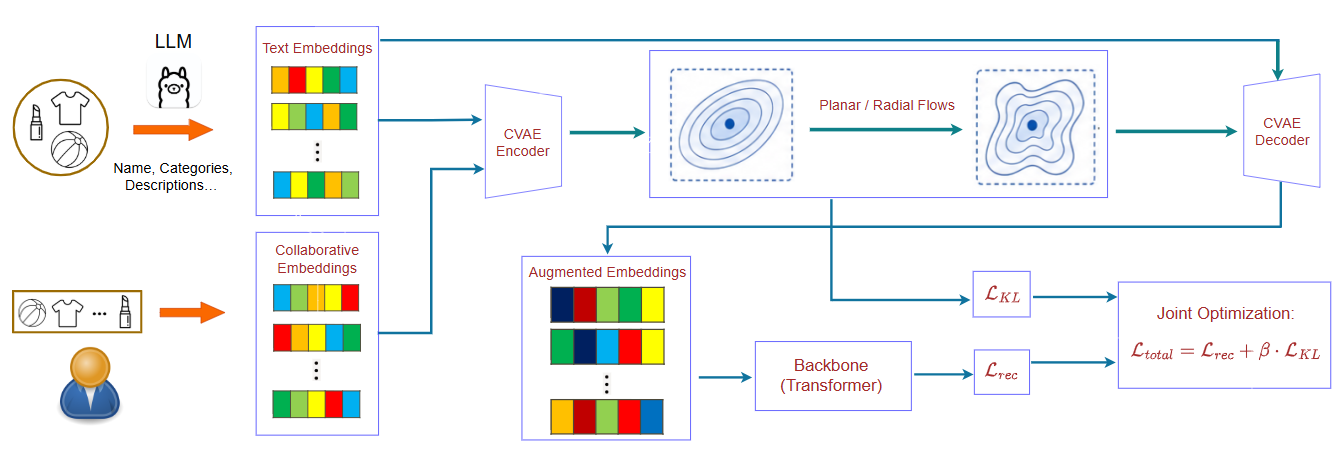}
    \caption{Overview of the proposed LSF-SR framework. The architecture uses LLMs to synthesize descriptive item prose offline and to extract semantic signals from text. Grounded in these signals, the conditional semantic fusion module, which incorporates normalizing flows to reshape the latent posterior, and the Transformer-based sequential backbone are jointly optimized end-to-end under a unified objective.}
    \label{fig:framework}
    \Description{Overview of LSF-SR. Offline LLM profiling produces item text that is encoded into semantic conditions. A flow-augmented conditional fusion module maps collaborative item embeddings into enhanced representations, which a Transformer backbone uses for next-item prediction.}
\end{figure*}

\section{Preliminaries}
\label{sec:preliminaries}
This section establishes the formal notation and background needed to describe the proposed framework.

\subsection{Problem Formulation}
\label{subsec:problem}
Let $\mathcal{U} = \{u_1, u_2, \dots, u_{|\mathcal{U}|}\}$ and $\mathcal{I} = \{i_1, i_2, \dots, i_{|\mathcal{I}|}\}$ denote the sets of users and items, respectively. For each user $u \in \mathcal{U}$, the historical interaction sequence is $S^u = (i_1, i_2, \dots, i_{t}, \dots, i_{|S^u|})$, where $i_t \in \mathcal{I}$ is the item interacted with at timestep $t$. The next-item prediction task is to develop a model that predicts the most likely next item, denoted $i_{|S^u|+1}$. The model estimates the probability $P(i_{|S^u|+1} = j \mid S^u)$ for all candidates $j \in \mathcal{I}$ and generates a ranked list.

\subsection{Sequential Recommendation Backbone}
\label{subsec:backbone}
We use a Transformer-based architecture as the foundation for our sequential recommendation model to capture evolving preferences in user-item interactions.
The backbone typically comprises three main components:

\begin{enumerate}
  \item \textbf{Embedding Layer}: Each item ID is mapped to a $d$-dimensional embedding $\mathbf{e}_i \in \mathbb{R}^d$. In LSF-SR, these embeddings are first passed through the conditional semantic fusion module to produce enhanced representations $\mathbf{e}'_i$. Learnable positional embeddings are then added to $\mathbf{e}'_i$ before the Transformer encoder to preserve temporal order.
  
  \item \textbf{Self-Attention Blocks}: Multiple layers of multi-head self-attention are used to adaptively aggregate information. For each head, the attention mechanism computes the relevance between items using Query ($Q$), Key ($K$), and Value ($V$) matrices derived from the sequence representations. Let $d_k$ denote the per-head key dimension.
  \begin{equation}
      \text{Attention}(Q, K, V) = \text{softmax}\left(\frac{QK^{\top}}{\sqrt{d_k}}\right)V
  \end{equation}
  
  \item \textbf{Point-wise Feed-Forward Network (FFN)}: After the attention layers, a two-layer FFN with non-linear activations is applied independently at each position to enhance the encoder's non-linear modeling capability.
\end{enumerate}

The backbone outputs a comprehensive sequence representation $\mathbf{h}_u \in \mathbb{R}^d$ that captures the user's latent intent and serves as the basis for the recommendation.

\section{Our Proposed LSF-SR}
\label{sec:method}
In this section, we present the technical details of our proposed \textbf{L}atent \textbf{S}emantic \textbf{F}usion for \textbf{S}equential \textbf{R}ecommendation via Flow-based Conditional Variational Autoencoders (\textbf{LSF-SR}). 
As shown in Figure~\ref{fig:framework}, the framework first uses LLMs to synthesize descriptive item prose and extract semantic signals from text.
The conditional semantic fusion module then applies normalizing flows to reshape the latent posterior and generate condition-aware item representations, thereby bridging the representation gap between collaborative signals and textual semantic conditions.
Finally, we integrate the enriched item representations into the Transformer-based sequential backbone to model dynamic user preferences end-to-end.

\subsection{Item Semantic Acquisition}
\label{subsec:acquisition}
We use a pre-trained LLM's reasoning capabilities to profile item semantics.
This process transforms fragmented item information into a cohesive semantic context, providing a strong foundation for representation learning.

\subsubsection{LLM-based Profiling}
Instead of passing raw metadata directly to the recommendation backbone, we recast item profiling as a generative task. 
For each item $i \in \mathcal{I}$, we assemble a detailed prompt $P_i$ that combines its available attributes $A_i$ with a representative subset of historical interaction sequences $H_i$. 
We then query the LLM to produce a descriptive text $T_i$ that captures the item's factual characteristics and hypothesized user intent:
\begin{equation}
    T_i = \text{LLM}(P_i)
\end{equation}
The generated description $T_i$ serves as a semantic intermediary, transforming sparse, discrete, and potentially incomplete attributes into a coherent textual representation. 
This enriched representation mitigates the limitations of static metadata and provides a more informative, context-aware input for downstream components.

\subsubsection{Semantic Representation}
To embed the descriptive text $T_i$ into a latent representation, we use a pre-trained semantic encoder that outputs a dense vector $\mathbf{c}_i \in \mathbb{R}^{d_s}$:
\begin{equation}
    \mathbf{c}_i = \text{Encoder}(T_i)
\end{equation}
where $d_s$ is the dimensionality of the semantic embedding space. 
To align these semantic vectors with the hidden space of the sequential recommender, we introduce a linear projection layer that yields the projected semantic vector $\hat{\mathbf{c}}_i \in \mathbb{R}^{d}$:
\begin{equation}
    \hat{\mathbf{c}}_i = \mathbf{W}_s \mathbf{c}_i + \mathbf{b}_s
\end{equation}
where $\mathbf{W}_s \in \mathbb{R}^{d \times d_s}$ and $\mathbf{b}_s \in \mathbb{R}^d$ are learnable parameters. 
The resulting projected embeddings capture high-level semantic signals and condition the subsequent fusion module, anchoring collaborative signals in rich semantic information.

\subsection{Conditional Semantic Fusion}
\label{subsec:fusion}
Integrating LLM knowledge into sequential recommendation systems requires precise alignment between collaborative and textual semantic spaces.
To address this challenge, we propose a conditional fusion module within the conditional variational autoencoder framework.
This module frames fusion as a stochastic inference task, capturing complex nonlinear interactions more effectively than deterministic mappings.

\subsubsection{Probabilistic Fusion Framework}
Let $\mathbf{e}_i \in \mathbb{R}^d$ denote the collaborative embedding of item $i$, and $\hat{\mathbf{c}}_i \in \mathbb{R}^d$ denote its corresponding semantic condition embedding. 
Our goal is to learn a latent variable \(\mathbf{z}\) that synthesizes these two representations. 
In the conditional variational autoencoder framework, the semantic condition embedding \(\hat{\mathbf{c}}_i\) serves as a conditioning variable that guides the transformation of the collaborative embedding. 
The CVAE infers an approximate posterior
\begin{equation}
    q_{\phi}(\mathbf{z} | \mathbf{e}_i, \hat{\mathbf{c}}_i)
\end{equation}
where \(\phi\) denotes the encoder parameters.
By modeling this distribution, the fusion module accounts for uncertainty when aligning item identities with their semantic context, rather than assuming a deterministic mapping between the two spaces.
Unlike a reconstruction autoencoder, the decoder generates enhanced embeddings $\mathbf{e}'_i$ for next-item prediction rather than reconstructing $\mathbf{e}_i$.

\subsubsection{Latent Encoding and Reparameterization}
The conditional encoder maps the joint input to a latent space by processing the concatenation of the collaborative embedding and the semantic condition:
\begin{equation}
    [\boldsymbol{\mu}_i, \log \boldsymbol{\sigma}_i^2] = \text{MLP}_{\text{enc}}([\mathbf{e}_i; \hat{\mathbf{c}}_i])
\end{equation}
where $\boldsymbol{\mu}_i, \boldsymbol{\sigma}_i \in \mathbb{R}^{d_z}$ denote the mean and standard deviation of the base Gaussian, and $d_z$ is the latent dimensionality.
To enable stable end-to-end optimization via stochastic sampling, we apply the reparameterization trick to obtain a base latent variable:
\begin{equation}
    \mathbf{z}_0 = \boldsymbol{\mu}_i + \boldsymbol{\epsilon} \odot \boldsymbol{\sigma}_i, \quad \boldsymbol{\epsilon} \sim \mathcal{N}(\mathbf{0}, \mathbf{I})
\end{equation}
This ensures that $\mathbf{z}_0$ captures synthesized features from both spaces while remaining differentiable with respect to the inputs.

\subsubsection{Condition-aware Embedding Generation}
The final stage uses a conditional decoder to generate the enhanced item representation $\mathbf{e}'_i$. 
After reparameterization, the base latent code $\mathbf{z}_0$ can be transformed via normalizing flows (Section~\ref{subsec:flows}) into $\mathbf{z}_K$.
The decoder conditions on this latent code and the semantic condition $\hat{\mathbf{c}}_i$:
\begin{equation}
    \mathbf{e}'_i = \text{MLP}_{\text{dec}}([\mathbf{z}_K; \hat{\mathbf{c}}_i])
\end{equation}
where $\mathbf{z}_K = \mathbf{z}_0$ when the flow depth is $K=0$.
By incorporating the semantic condition into both the encoder and the decoder, LSF-SR establishes a unified latent manifold in which collaborative and semantic features are closely aligned.
This design reduces the mismatch that often arises when collaborative and semantic features are combined via concatenation or other deterministic fusion operators.

\subsection{Expressive Posterior with Normalizing Flows}
\label{subsec:flows}
Although a Gaussian posterior is a fundamental assumption in the conditional variational autoencoder paradigm, the complex, nonlinear relationships between collaborative signals and semantic conditions often require a more flexible posterior.
To address this, we incorporate normalizing flows to make the latent space more expressive.
This approach transforms a simple initial density into a complex posterior distribution through a series of invertible mappings, enabling a more accurate representation of item characteristics on the latent manifold.

\subsubsection{Probability Transformation via Flows}
The transformation process begins with an initial latent variable $\mathbf{z}_0$ sampled from the base Gaussian distribution.
We then apply a chain of $K$ invertible transformations $f_1, f_2, \dots, f_K$ to obtain the final latent variable $\mathbf{z}_K = f_K \circ \dots \circ f_1(\mathbf{z}_0)$. 
By the change-of-variables formula, the log-probability of the final variable is:
\begin{equation}
    \ln q_K(\mathbf{z}_K) = \ln q_0(\mathbf{z}_0) - \sum_{k=1}^K \ln \left| \det \frac{\partial f_k}{\partial \mathbf{z}_{k-1}} \right|
\end{equation}
The second term, the sum of the log-determinants of the Jacobians, captures the volume change induced by each transformation. 
This warping mechanism enables LSF-SR to refine the latent space so that it more accurately reflects the underlying item semantics, especially for complex, non-Gaussian feature distributions.

\subsubsection{Planar and Radial Flow Stacks}
We implement two types of flow layers to adapt to varying data characteristics across domains. 
Planar flows apply transformations along specific directions via hyperplanes, which are particularly effective at capturing nonlinear boundaries between distinct clusters in the semantic manifold. 
The transformation for each layer is defined as:
\begin{equation}
    f(\mathbf{z}) = \mathbf{z} + \mathbf{u}\,\tanh(\mathbf{w}^T \mathbf{z} + b)
\end{equation}
where $\mathbf{w}, \mathbf{u} \in \mathbb{R}^{d_z}$ and $b \in \mathbb{R}$ are learnable parameters. 
In contrast, radial flows model localized semantic clusters by applying transformations centered on a reference point $\mathbf{z}_{\text{ref}}$:
\begin{equation}
    f(\mathbf{z}) = \mathbf{z} + \beta_r \cdot \frac{1}{\alpha + \|\mathbf{z} - \mathbf{z}_{\text{ref}}\|_2} \cdot (\mathbf{z} - \mathbf{z}_{\text{ref}})
\end{equation}
where $\alpha > 0$, $\beta_r \in \mathbb{R}$, and $\mathbf{z}_{\text{ref}} \in \mathbb{R}^{d_z}$ are parameters optimized during training.
Stacking multiple layers of these flows allows the model to refine the latent posterior adaptively, ensuring high-fidelity fusion of collaborative signals and semantic conditions.

\subsection{Training Objectives and KL Annealing}
\label{subsec:training}
LSF-SR training is governed by a joint objective function that optimizes both recommendation accuracy and the structural regularity of the fused latent space.
We introduce a dynamic KL annealing mechanism to address posterior collapse, a known issue in which the model effectively ignores the latent semantic variables.
This adaptive strategy promotes stable convergence and ensures that collaborative signals are grounded in the semantic conditions.

\subsubsection{Recommendation Loss}
The primary goal of LSF-SR is to predict the next item in a sequence.
For each training instance, the model maps items in the input sequence to collaborative embeddings, fuses them into enhanced representations, adds positional embeddings, and encodes them with the Transformer backbone to obtain $\mathbf{h}_u$.
Enhanced embeddings for the full item catalog, denoted $\mathbf{E}' = \{\mathbf{e}'_1, \dots, \mathbf{e}'_{|\mathcal{I}|}\}$, are obtained from the fusion module in inference mode.
The next-item probability is computed via a softmax over the full catalog:
\begin{equation}
    \hat{y}_{uj} = \frac{\exp(\mathbf{h}_u^{\top} \mathbf{e}'_j)}{\sum_{k \in \mathcal{I}} \exp(\mathbf{h}_u^{\top} \mathbf{e}'_k)}
\end{equation}
The recommendation loss $\mathcal{L}_{rec}$ is the negative log-likelihood of the ground-truth next item $i^+$ for each training instance:
\begin{equation}
    \mathcal{L}_{rec} = -\sum_{(S^u, i^+)} \ln \hat{y}_{u, i^+}
\end{equation}
During training, KL regularization is computed on the stochastic fusion path for sequence items, whereas full-catalog logits for $\mathcal{L}_{rec}$ use the deterministic inference path ($\mathbf{z}_K=\text{flow}(\boldsymbol{\mu}_i)$ without sampling).

\subsubsection{KL Divergence in Conditional Fusion Module}
In our conditional fusion module, the variational objective minimizes the KL divergence between the approximate posterior and the prior.
When incorporating normalizing flows, the KL term for each item $i$ is expressed as an expectation over the base distribution $q_0(\mathbf{z}_0)$:
\begin{equation}
    \mathcal{L}_{KL} = \mathbb{E}_{q_0(\mathbf{z}_0)} \left[ \ln q_0(\mathbf{z}_0) - \sum_{k=1}^K \ln \left| \det \frac{\partial f_k}{\partial \mathbf{z}_{k-1}} \right| - \ln p(\mathbf{z}_K) \right]
\end{equation}
where $\mathbf{z}_K$ is the latent variable after $K$ successive flow transformations.
We use a standard isotropic Gaussian prior $p(\mathbf{z}_K)=\mathcal{N}(\mathbf{0}, \mathbf{I})$.
This term regularizes the approximate posterior toward the prior while preserving informative fused representations.

\subsubsection{Dynamic KL Annealing Strategies}
The overall training objective is a weighted combination of the aforementioned losses:
\begin{equation}
    \mathcal{L}_{total} = \mathcal{L}_{rec} + \beta \cdot \mathcal{L}_{KL}
\end{equation}
where $\beta$ serves as a dynamic regularization coefficient. 
To prevent the optimization from converging to suboptimal local minima, we implement three annealing strategies for $\beta$ as a function of the training epoch $e$:
\begin{itemize}[leftmargin=*, topsep=1pt, itemsep=1pt, parsep=0pt]
    \item \textbf{Linear Annealing}: $\beta$ increases linearly from a starting value to its maximum, providing steady latent-space regularization.
    \item \textbf{Sigmoid Annealing}: $\beta$ follows a sigmoid curve, enabling a smooth, non-linear transition that prioritizes early recommendation accuracy before enforcing strict semantic alignment.
    \item \textbf{Cyclical Annealing}: $\beta$ follows a periodic sawtooth pattern, allowing the model to alternate between recommendation and regularization phases, refining representations multiple times during training.
\end{itemize}

\begin{figure*}[htpb]
    \centering    
    \includegraphics[width=1\textwidth]{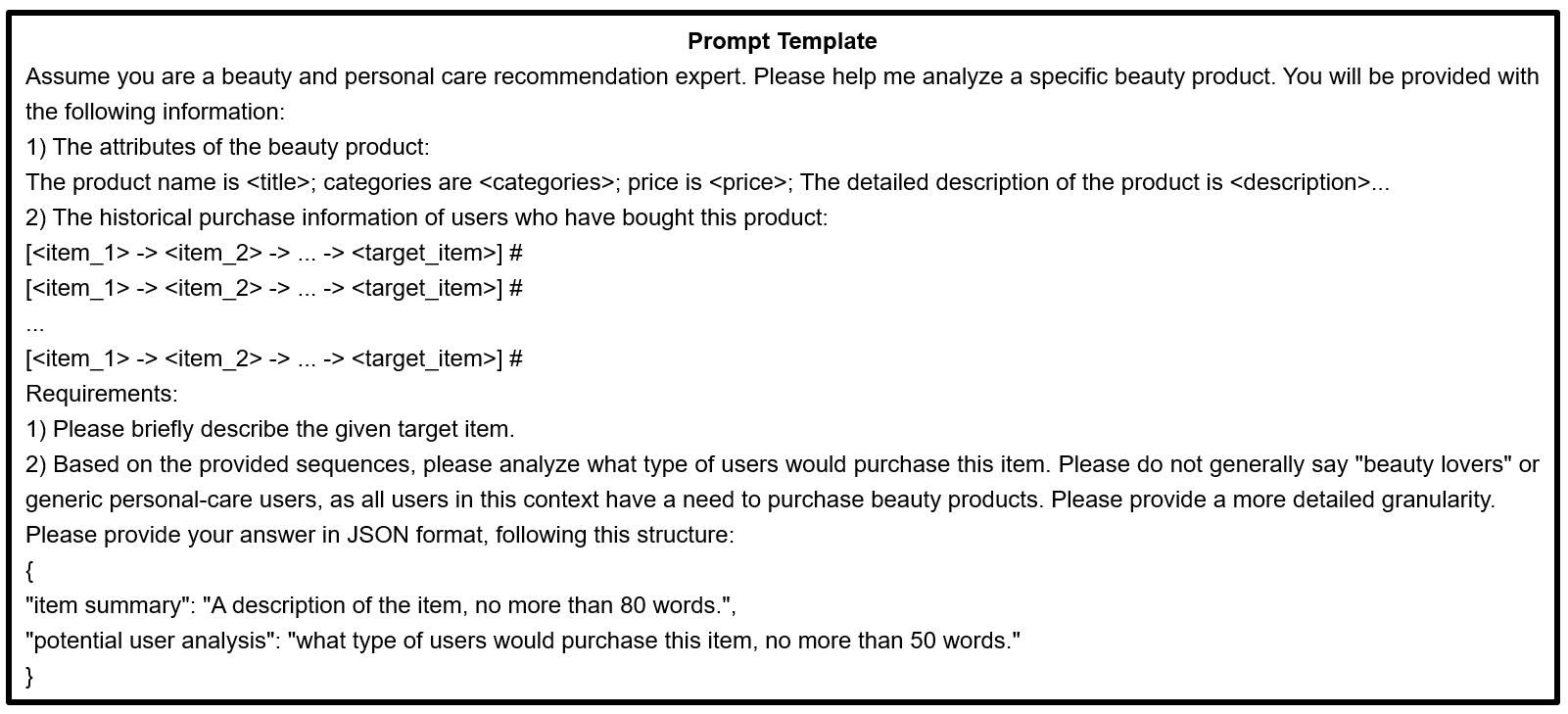} 
    \caption{The structured prompt template used to generate semantic textual profiles for the Amazon Beauty dataset.}
    \label{fig:prompt_template}
    \Description{A structured prompt that asks an LLM to write an item profile from Beauty metadata and a sample of historical interactions, covering product attributes and hypothesized user intent.}
\end{figure*}

\section{Experiments}
\label{sec:exp}
In this section, we present a comprehensive empirical evaluation of \textbf{LSF-SR} across five public benchmark datasets.
We conduct all experiments using PyTorch 1.13 (CUDA 11.7) and Python 3.8.19 on a Linux workstation with an Intel Xeon Gold 6140 CPU (2.30 GHz) and a single NVIDIA GeForce RTX 4090 GPU.

\subsection{Datasets}
We evaluate the performance of our LSF-SR framework on five widely used public datasets: four domains from the Amazon review dataset (Beauty, Office, Sports, and Toys) and the Yelp open dataset \cite{amazon, yelp}. 
The Amazon datasets span diverse consumer product categories and include user transaction histories and item metadata. 
The Yelp dataset is a well-known resource for business recommendations and captures user reviews of local establishments. 
Consistent with the literature \cite{diff, sracl}, we filter Yelp interactions to 2019 and apply standard 5-core filtering across all datasets. 
Table~\ref{tab:datasets} summarizes the statistics of the preprocessed datasets.

\begin{table}[t] 
\caption{Data statistics after preprocessing. Avg. Length indicates the average number of interactions per user.}
\label{tab:datasets}
\centering
\begin{tabular}{l|ccccc}
\toprule
Dataset & Beauty & Office & Sports & Toys & Yelp \\
\midrule
\# Users & 22,363 & 4,905 & 35,598 & 19,412 & 19,936 \\
\# Items & 12,101 & 2,420 & 18,357 & 11,924 & 14,587 \\
\# Interactions & 198,502 & 53,258 & 296,337 & 167,597 & 207,952 \\
Avg. Length & 8.9 & 10.9 & 8.3 & 8.6 & 10.4 \\
Sparsity & 99.93\% & 99.55\% & 99.95\% & 99.93\% & 99.93\% \\
\bottomrule
\end{tabular}
\end{table}

\subsection{Evaluation Metrics} 
To assess the recommendation model's performance, we adopt the leave-one-out protocol, commonly used in sequential recommendation tasks. 
For each user interaction sequence, we designate the last interaction as the test set, the penultimate interaction as the validation set, and use all remaining historical interactions for training. 
During evaluation, we follow the full-ranking protocol, ranking the ground-truth next item against the entire item space to ensure an unbiased, thorough assessment \cite{krichene2020sampled}. 
To quantify recommendation accuracy and ranking quality, we employ two standard information retrieval metrics: Recall@$K$ and Normalized Discounted Cumulative Gain (NDCG@$K$), with $K \in \{10, 20\}$.

\subsection{Baseline Methods} 
To demonstrate LSF-SR's effectiveness, we compare it with several competitive baselines across two main paradigms: Conventional Behavior-Centric Sequential Recommendation and LLM-augmented Sequential Recommendation. 
The behavior-centric group includes the foundational self-attention architecture (SASRec) \cite{sasrec}, along with state-of-the-art models from three sub-paradigms: (1) contrastive learning methods (DuoRec, ICLRec, MCLRec, ICSRec, and FENRec) \cite{duorec, iclrec, mclrec, icsrec, fenrec}; (2) frequency-enhanced filters (FMLP-Rec, FEARec, and BSARec) \cite{fmlprec, fearec, bsarec}; and (3) side-information-integrated sequential recommendation, represented by the dual side-information filtering and fusion framework (DIFF) \cite{diff}. 
In contrast, the LLM-augmented domain includes prominent representative methods (LRD, LLM-ESR, and SRA-CL) \cite{lrd, llmesr, sracl} that use LLM-generated textual semantics to contextualize sequential behavior patterns.

\subsection{Implementation Details} 
In our model, we set the item ID embedding dimension and the Transformer hidden size to 64.
The backbone uses two self-attention layers with two attention heads, applies dropout of $0.3$ to both hidden units and attention weights, and limits the sequence length to 20.
We use a learning rate of $0.001$ ($0.0001$ on Yelp).
The conditional semantic fusion module uses a 32-dimensional latent space ($d_z=32$), and both the conditional encoder and decoder are implemented as two-layer MLPs with layer normalization and GELU activations. 
To enhance posterior expressiveness, we use planar flows with $K=2$ on Beauty and Sports, planar flows with $K=3$ on Toys and Yelp, and radial flows with $K=3$ on Office.
We set $\beta_{\mathrm{end}}$ to 0.1 on Beauty, Sports, and Yelp, 0.15 on Toys, and 0.2 on Office.
We adopt cyclical KL annealing by default, raising $\beta$ from $\beta_{\mathrm{start}}=0.01$ ($0.001$ on Yelp) to the selected $\beta_{\mathrm{end}}$ over cycles of 30 epochs (40 on Yelp). 
For all baseline models, we adhere strictly to the hyperparameter configurations reported in their original literature. 
If specific details are not provided, we carefully tune the parameters to achieve optimal performance. 
To ensure a rigorous and fair comparison, the baseline implementations are standardized under two strict protocols: (1) for both side-information-integrated and LLM-augmented approaches, we consistently use the same raw data fields to maintain historical input consistency; and (2) for all LLM-augmented models, we standardize the text processing pipeline by using Meta LLaMA 3.1-8B-Instruct as the generative language model and supervised SimCSE (RoBERTa-large; $d_s=1024$) as the textual embedding extractor \cite{llama3herd2024, gao2021simcse}.
Figure~\ref{fig:prompt_template} illustrates the representative prompt template developed for the Amazon Beauty dataset.

\begin{table*}[t]
\centering
\caption{Overall performance comparison across five benchmark datasets. The best baseline results are underlined, and our proposed method and its improvements are highlighted in \textbf{bold}. $^*$ indicates a statistically significant improvement over the best baseline ($p < 0.05$) using two-tailed paired $t$-tests.}
\label{tab:performance_comparison_grouped_lines}
\resizebox{\textwidth}{!}{
\begin{tabular}{ll|cccccccccc|ccc|c|c}
\toprule
\multirow{2}{*}{Dataset} & \multirow{2}{*}{Metric} & \multicolumn{10}{c|}{Conventional Behavior-Centric SR Baselines} & \multicolumn{3}{c|}{LLM-Augmented SR Baselines} & (Ours) & \multirow{2}{*}{Improv.} \\
\cmidrule(lr){3-12} \cmidrule(lr){13-15}
& & SASRec & ICLRec & DuoRec & MCLRec & ICSRec & FENRec & FMLP-Rec & FEARec & BSARec & DIFF & LRD & LLM-ESR & SRA-CL & LSF-SR & \\
\midrule
\multirow{4}{*}{Beauty} 
& R@10 & 0.0798 & 0.0846 & 0.0828 & 0.0849 & 0.0857 & 0.0898 & 0.0885 & 0.0880 & 0.0872 & 0.0845 & 0.0760 & 0.0663 & \underline{0.0919} & \textbf{0.0969}\textsuperscript{*} & \textbf{5.51\%} \\
& N@10 & 0.0404 & 0.0436 & 0.0437 & 0.0439 & 0.0445 & 0.0466 & 0.0444 & 0.0454 & 0.0443 & \underline{0.0481} & 0.0364 & 0.0304 & 0.0466 & \textbf{0.0520}\textsuperscript{*} & \textbf{8.11\%} \\
& R@20 & 0.1140 & 0.1195 & 0.1190 & 0.1215 & 0.1213 & 0.1286 & 0.1272 & 0.1242 & 0.1265 & 0.1209 & 0.1149 & 0.1019 & \underline{0.1323} & \textbf{0.1406}\textsuperscript{*} & \textbf{6.26\%} \\
& N@20 & 0.0490 & 0.0524 & 0.0527 & 0.0532 & 0.0534 & 0.0564 & 0.0542 & 0.0545 & 0.0542 & \underline{0.0572} & 0.0462 & 0.0394 & 0.0568 & \textbf{0.0629}\textsuperscript{*} & \textbf{10.00\%} \\
\midrule
\multirow{4}{*}{Office} 
& R@10 & 0.0968 & 0.0875 & 0.0979 & 0.1056 & 0.1023 & 0.1063 & 0.1043 & 0.1032 & 0.1057 & \underline{0.1133} & 0.1099 & 0.0924 & 0.1074 & \textbf{0.1281}\textsuperscript{*} & \textbf{12.99\%} \\
& N@10 & 0.0484 & 0.0403 & 0.0507 & 0.0545 & 0.0524 & 0.0567 & 0.0518 & 0.0529 & 0.0528 & \underline{0.0586} & 0.0490 & 0.0432 & 0.0564 & \textbf{0.0667}\textsuperscript{*} & \textbf{13.85\%} \\
& R@20 & 0.1487 & 0.1411 & 0.1514 & 0.1635 & 0.1601 & 0.1689 & 0.1627 & 0.1624 & 0.1648 & \underline{0.1767} & 0.1751 & 0.1439 & 0.1689 & \textbf{0.1925}\textsuperscript{*} & \textbf{8.97\%} \\
& N@20 & 0.0615 & 0.0537 & 0.0642 & 0.0690 & 0.0669 & 0.0724 & 0.0664 & 0.0678 & 0.0676 & \underline{0.0745} & 0.0654 & 0.0561 & 0.0719 & \textbf{0.0830}\textsuperscript{*} & \textbf{11.30\%} \\
\midrule
\multirow{4}{*}{Sports} 
& R@10 & 0.0470 & 0.0506 & 0.0475 & 0.0480 & 0.0501 & \underline{0.0547} & 0.0545 & 0.0514 & 0.0535 & 0.0507 & 0.0407 & 0.0374 & 0.0529 & \textbf{0.0614}\textsuperscript{*} & \textbf{12.26\%} \\
& N@10 & 0.0235 & 0.0250 & 0.0244 & 0.0248 & 0.0248 & \underline{0.0284} & 0.0259 & 0.0250 & 0.0254 & 0.0271 & 0.0184 & 0.0174 & 0.0272 & \textbf{0.0324}\textsuperscript{*} & \textbf{14.14\%} \\
& R@20 & 0.0692 & 0.0736 & 0.0703 & 0.0721 & 0.0727 & \underline{0.0812} & 0.0805 & 0.0758 & 0.0791 & 0.0762 & 0.0648 & 0.0575 & 0.0801 & \textbf{0.0917}\textsuperscript{*} & \textbf{12.98\%} \\
& N@20 & 0.0291 & 0.0308 & 0.0302 & 0.0308 & 0.0305 & \underline{0.0351} & 0.0324 & 0.0312 & 0.0319 & 0.0335 & 0.0245 & 0.0225 & 0.0341 & \textbf{0.0401}\textsuperscript{*} & \textbf{14.13\%} \\
\midrule
\multirow{4}{*}{Toys} 
& R@10 & 0.0828 & 0.0962 & 0.0936 & 0.0905 & 0.0963 & \underline{0.1000} & 0.0946 & 0.0968 & 0.0959 & 0.0913 & 0.0804 & 0.0745 & 0.0960 & \textbf{0.1086}\textsuperscript{*} & \textbf{8.58\%} \\
& N@10 & 0.0411 & \underline{0.0521} & 0.0476 & 0.0462 & 0.0500 & 0.0509 & 0.0480 & 0.0498 & 0.0480 & 0.0484 & 0.0362 & 0.0344 & 0.0488 & \textbf{0.0570}\textsuperscript{*} & \textbf{9.49\%} \\
& R@20 & 0.1148 & 0.1318 & 0.1281 & 0.1258 & 0.1302 & \underline{0.1380} & 0.1331 & 0.1308 & 0.1335 & 0.1300 & 0.1233 & 0.1078 & 0.1356 & \textbf{0.1545}\textsuperscript{*} & \textbf{11.95\%} \\
& N@20 & 0.0492 & \underline{0.0610} & 0.0563 & 0.0551 & 0.0585 & 0.0604 & 0.0577 & 0.0583 & 0.0576 & 0.0581 & 0.0470 & 0.0428 & 0.0588 & \textbf{0.0686}\textsuperscript{*} & \textbf{12.32\%} \\
\midrule
\multirow{4}{*}{Yelp} 
& R@10 & 0.0699 & 0.0760 & 0.0752 & 0.0737 & 0.0749 & \underline{0.0804} & 0.0800 & 0.0742 & 0.0785 & 0.0781 & 0.0686 & 0.0595 & 0.0777 & \textbf{0.0846}\textsuperscript{*} & \textbf{5.30\%} \\
& N@10 & 0.0376 & 0.0404 & 0.0398 & 0.0394 & 0.0398 & 0.0416 & \underline{0.0424} & 0.0391 & 0.0419 & 0.0408 & 0.0324 & 0.0308 & 0.0403 & \textbf{0.0436}\textsuperscript{*} & \textbf{2.78\%} \\
& R@20 & 0.1099 & 0.1187 & 0.1176 & 0.1155 & 0.1187 & \underline{0.1283} & 0.1244 & 0.1155 & 0.1230 & 0.1250 & 0.1119 & 0.0936 & 0.1247 & \textbf{0.1363}\textsuperscript{*} & \textbf{6.24\%} \\
& N@20 & 0.0477 & 0.0511 & 0.0504 & 0.0499 & 0.0508 & \underline{0.0537} & 0.0535 & 0.0495 & 0.0530 & 0.0525 & 0.0432 & 0.0393 & 0.0522 & \textbf{0.0566}\textsuperscript{*} & \textbf{5.40\%} \\
\bottomrule
\end{tabular}
}
\end{table*}

\subsection{Experimental Results}
We first compare LSF-SR with state-of-the-art methods, then ablate core components and KL annealing schedules, and finally analyze hyperparameter sensitivity and cold-start case studies.

\subsubsection{Comparison with SOTA}
\label{sec:overall_performance}
Table~\ref{tab:performance_comparison_grouped_lines} presents a comprehensive performance comparison of our proposed LSF-SR against several competitive baselines across five benchmark datasets.
To ensure statistical validity, we ran each experiment with five random seeds and report the average results.
We validated these results with two-tailed paired $t$-tests, confirming that LSF-SR's performance gains are statistically significant ($p < 0.05$).   

Empirically, LSF-SR consistently and significantly outperforms all baselines across datasets and metrics.
Under full-ranking evaluation, conventional behavior-centric models remain competitive and often outperform LLM-augmented baselines.
LLM-ESR, in particular, lags behind strong behavior-centric methods on all five datasets; LRD is similarly weak on most datasets, though it remains competitive on Office.
These results suggest that methods developed primarily under sampled-negative protocols may not transfer as well to exhaustive ranking.
SRA-CL narrows this gap through LLM-based semantic retrieval and contrastive learning, yet its decoupled retrieval and alignment still limit bridging between textual semantic and collaborative spaces.
Overall, LSF-SR's superiority over both conventional and LLM-augmented models indicates that probabilistic fusion better exploits complementary behavioral and textual semantic signals.

\subsubsection{Ablation Study}
\label{subsec:ablation}
This subsection evaluates each core component of LSF-SR and examines how different optimization schedules affect performance.
Table~\ref{tab:ablation_study} summarizes the results, highlighting performance drops across ablated modules and across KL annealing schemes.

We examine three architectural variants, each designed to isolate a key component and evaluate its impact. 
In the \textbf{w/o LLM} setting, we skip the generative profiling stage and directly transform raw metadata prompt templates into text embeddings, which then serve as semantic conditions. 
Consistently weaker results across all datasets indicate that unprocessed, fragmented metadata fails to provide effective semantic guidance for alignment. 
This highlights the crucial role of the LLM’s semantic synthesis in generating rich, expressive item profiles. 
Notably, the performance drop is smallest on the Yelp dataset, which has metadata constraints. 
Unlike the Amazon datasets, raw Yelp metadata lacks extensive textual features, such as detailed descriptions, limiting the quality of the generated prose and the associated incremental gains. 
This discrepancy shows that LSF-SR's performance advantage is intrinsically linked to the input metadata's contextual richness.

We also explore the latent space's structural properties. 
In the \textbf{w/o Flows} setting, we remove the normalizing flows by setting the flow depth to $K=0$, which constrains the approximate posterior to a learned diagonal Gaussian. 
The resulting performance decline underscores the importance of invertible, nonlinear transformations for reshaping the latent space into an expressive manifold. 
Notably, this reduction remains relatively minor for Sports; on Yelp, Recall@20 drops slightly while NDCG@20 is unchanged relative to the full model. 
We hypothesize that the larger item scales and intrinsic sparsity in these domains allow their underlying geometric co-occurrence patterns to resemble a normal distribution closely. 
As a result, a simple Gaussian posterior provides adequate representational capacity.
By contrast, datasets with more complex collaborative--semantic dependencies benefit more from the additional geometric flexibility of the flow stacks.  

Finally, in the \textbf{w/o KL loss} setup, we eliminate latent-space regularization by explicitly setting the regularization weight $\beta = 0$. 
Among the ablations, removing the KL divergence constraint produces the largest performance drop. 
This clearly shows that our conditional fusion module relies on latent-space regularization; without the KL penalty, the posterior is unconstrained and can overfit idiosyncratic fusion patterns, ultimately compromising recommendation quality.

\begin{table*}[t]
\centering
\caption{An ablation study of model components and KL annealing strategies for LSF-SR. The right panel compares a fixed KL penalty with three dynamic annealing schedules: Linear, Sigmoid, and Cyclical. We adopt cyclical annealing as the default strategy.}
\label{tab:ablation_study}
\resizebox{\textwidth}{!}{
\begin{tabular}{ll|ccc|cccc}
\toprule
\multirow{2}{*}{Dataset} & \multirow{2}{*}{Metric} & \multicolumn{3}{c|}{Ablated Components} & \multicolumn{4}{c}{KL Annealing Strategies} \\
\cmidrule(lr){3-5} \cmidrule(lr){6-9}
& & w/o LLM & w/o Flows & w/o KL loss & Fixed & Linear & Sigmoid & Cyclical \\
\midrule
\multirow{2}{*}{Beauty} 
& R@20 & 0.1362 $\pm$ 0.0025 & 0.1384 $\pm$ 0.0017 & 0.1234 $\pm$ 0.0018 & 0.1315 $\pm$ 0.0012 & 0.1387 $\pm$ 0.0017 & 0.1373 $\pm$ 0.0023 & 0.1406 $\pm$ 0.0013 \\
& N@20 & 0.0606 $\pm$ 0.0010 & 0.0622 $\pm$ 0.0007 & 0.0563 $\pm$ 0.0007 & 0.0597 $\pm$ 0.0007 & 0.0619 $\pm$ 0.0012 & 0.0615 $\pm$ 0.0005 & 0.0629 $\pm$ 0.0007 \\
\midrule
\multirow{2}{*}{Office} 
& R@20 & 0.1806 $\pm$ 0.0042 & 0.1896 $\pm$ 0.0056 & 0.1750 $\pm$ 0.0039 & 0.1854 $\pm$ 0.0039 & 0.1855 $\pm$ 0.0030 & 0.1857 $\pm$ 0.0054 & 0.1925 $\pm$ 0.0028 \\
& N@20 & 0.0763 $\pm$ 0.0022 & 0.0815 $\pm$ 0.0028 & 0.0743 $\pm$ 0.0028 & 0.0793 $\pm$ 0.0023 & 0.0797 $\pm$ 0.0020 & 0.0793 $\pm$ 0.0024 & 0.0830 $\pm$ 0.0009 \\
\midrule
\multirow{2}{*}{Sports} 
& R@20 & 0.0877 $\pm$ 0.0006 & 0.0911 $\pm$ 0.0012 & 0.0813 $\pm$ 0.0017 & 0.0880 $\pm$ 0.0013 & 0.0887 $\pm$ 0.0012 & 0.0892 $\pm$ 0.0012 & 0.0917 $\pm$ 0.0008 \\
& N@20 & 0.0378 $\pm$ 0.0003 & 0.0397 $\pm$ 0.0003 & 0.0348 $\pm$ 0.0007 & 0.0383 $\pm$ 0.0004 & 0.0387 $\pm$ 0.0008 & 0.0388 $\pm$ 0.0004 & 0.0401 $\pm$ 0.0003 \\
\midrule
\multirow{2}{*}{Toys} 
& R@20 & 0.1450 $\pm$ 0.0023 & 0.1514 $\pm$ 0.0024 & 0.1288 $\pm$ 0.0012 & 0.1474 $\pm$ 0.0006 & 0.1549 $\pm$ 0.0017 & 0.1546 $\pm$ 0.0024 & 0.1545 $\pm$ 0.0010 \\
& N@20 & 0.0642 $\pm$ 0.0008 & 0.0677 $\pm$ 0.0009 & 0.0602 $\pm$ 0.0005 & 0.0654 $\pm$ 0.0004 & 0.0678 $\pm$ 0.0014 & 0.0681 $\pm$ 0.0009 & 0.0686 $\pm$ 0.0007 \\
\midrule
\multirow{2}{*}{Yelp} 
& R@20 & 0.1349 $\pm$ 0.0019 & 0.1359 $\pm$ 0.0020 & 0.1239 $\pm$ 0.0023 & 0.1303 $\pm$ 0.0020 & 0.1356 $\pm$ 0.0017 & 0.1342 $\pm$ 0.0025 & 0.1363 $\pm$ 0.0009 \\
& N@20 & 0.0557 $\pm$ 0.0007 & 0.0566 $\pm$ 0.0006 & 0.0524 $\pm$ 0.0007 & 0.0545 $\pm$ 0.0006 & 0.0562 $\pm$ 0.0006 & 0.0559 $\pm$ 0.0013 & 0.0566 $\pm$ 0.0005 \\
\bottomrule
\end{tabular}
}
\end{table*}

Beyond architectural components, Table~\ref{tab:ablation_study} compares a fixed $\beta$ penalty with three dynamic scheduling methods. 
A constant coefficient yields suboptimal performance because a rigid regularization penalty early in training severely constrains recommendation optimization and hinders effective semantic alignment.
In contrast, progressive scheduling techniques mitigate this issue by enabling a smoother transition between behavior modeling and semantic alignment.
Ultimately, the cyclical annealing strategy achieves the best or near-best NDCG@20 on all five datasets and the best or comparable Recall@20 on four of five datasets.
By periodically relaxing and tightening the KL constraints, this schedule gives the optimization process multiple opportunities to escape suboptimal local minima and adaptively refine the latent manifold geometry.

\begin{figure}[t]
    \centering
    \includegraphics[width=\columnwidth]{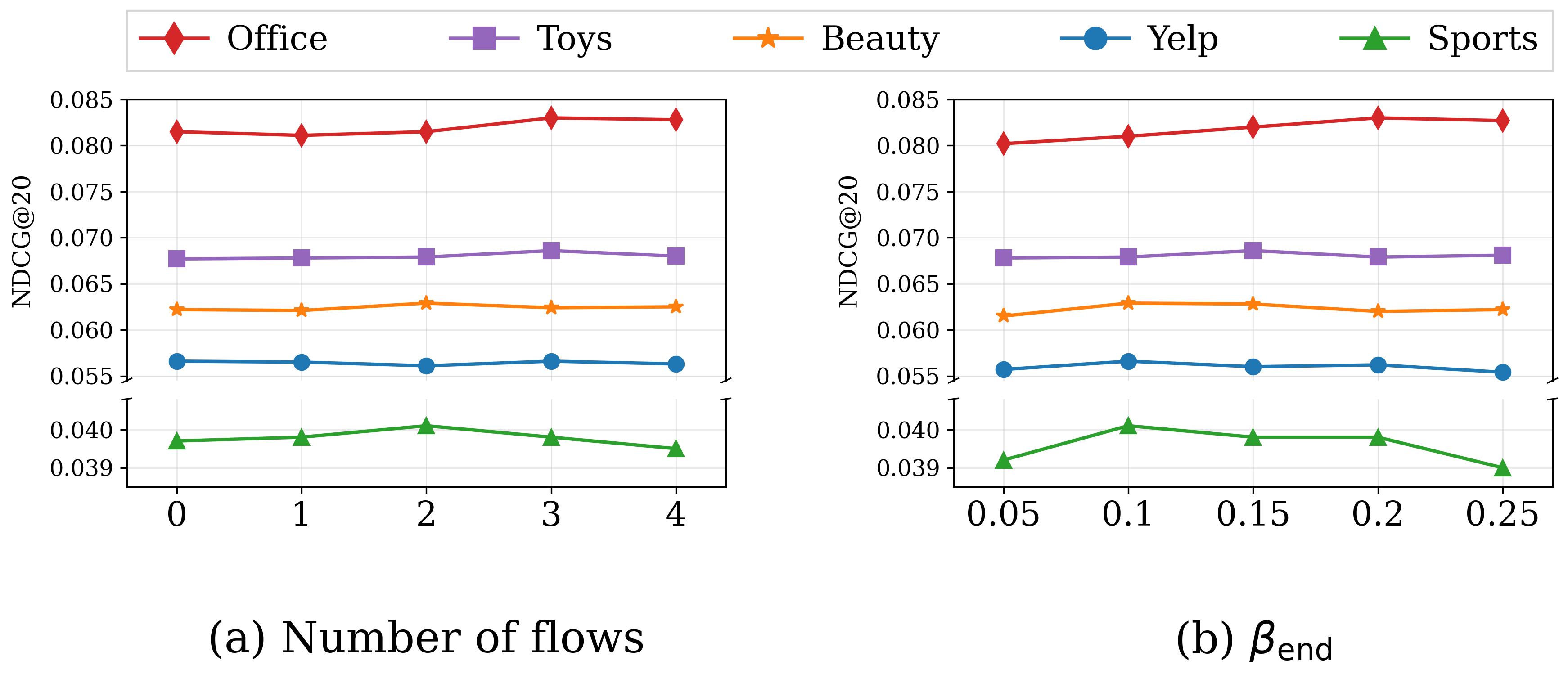}
    \Description{Line plots of NDCG@20 versus the number of flow layers $K$ (left) and the maximum KL coefficient $\beta_{\mathrm{end}}$ (right) on five datasets. Performance typically rises and then slightly declines as each parameter increases.}
    \caption{Analysis of hyperparameter sensitivity for LSF-SR across five datasets. The left panel shows the effect of the number of normalizing flows, while the right panel displays the impact of the maximum KL regularization coefficient ($\beta_{\mathrm{end}}$) on NDCG@20.}
    \label{fig:hyperparam_study}
\end{figure}



\subsubsection{Hyperparameter Study}
This subsection evaluates LSF-SR's sensitivity and stability with respect to two critical hyperparameters: the number of normalizing flows ($K$) and the maximum KL regularization coefficient ($\beta_{\mathrm{end}}$). 
These parameters control the flexibility of the latent-space transformation and the strength of the regularization term, respectively. 
As shown in Figure~\ref{fig:hyperparam_study}, the model's performance generally increases initially and then declines slightly as both parameters increase.

Regarding flow depth, setting $K=0$ constrains the latent manifold to a learned diagonal Gaussian posterior, inherently limiting its ability to capture complex patterns.
Increasing $K$ significantly improves the expressiveness of the normalizing flows, achieving peak performance by effectively aligning collaborative and textual semantic embeddings.
However, if $K$ becomes excessively large, it over-parameterizes the model, complicates optimization, and moderately reduces recommendation quality.

The impact of $\beta_{\mathrm{end}}$ presents a similar trade-off, balancing the main recommendation objective against semantic regularization. 
A small $\beta_{\mathrm{end}}$ provides insufficient constraints for effective semantic integration, whereas an appropriately scaled coefficient encourages the model to embed behavioral patterns in a rich semantic context. 
Conversely, if $\beta_{\mathrm{end}}$ exceeds the optimal threshold, the stringent KL penalty dominates the joint loss function. 
This over-regularization can suppress behavior-centric signals too much, ultimately compromising recommendation quality. 
Empirical evidence shows that LSF-SR achieves optimal performance by striking a delicate balance between behavioral discriminability and semantic regularization, a pattern consistently observed when $K$ is set to 2 or 3 and $\beta_{\mathrm{end}}$ falls between 0.1 and 0.2.

\subsection{Case Study: Cold Scenarios}
\begin{figure}[t]
    \centering
    \includegraphics[width=\columnwidth]{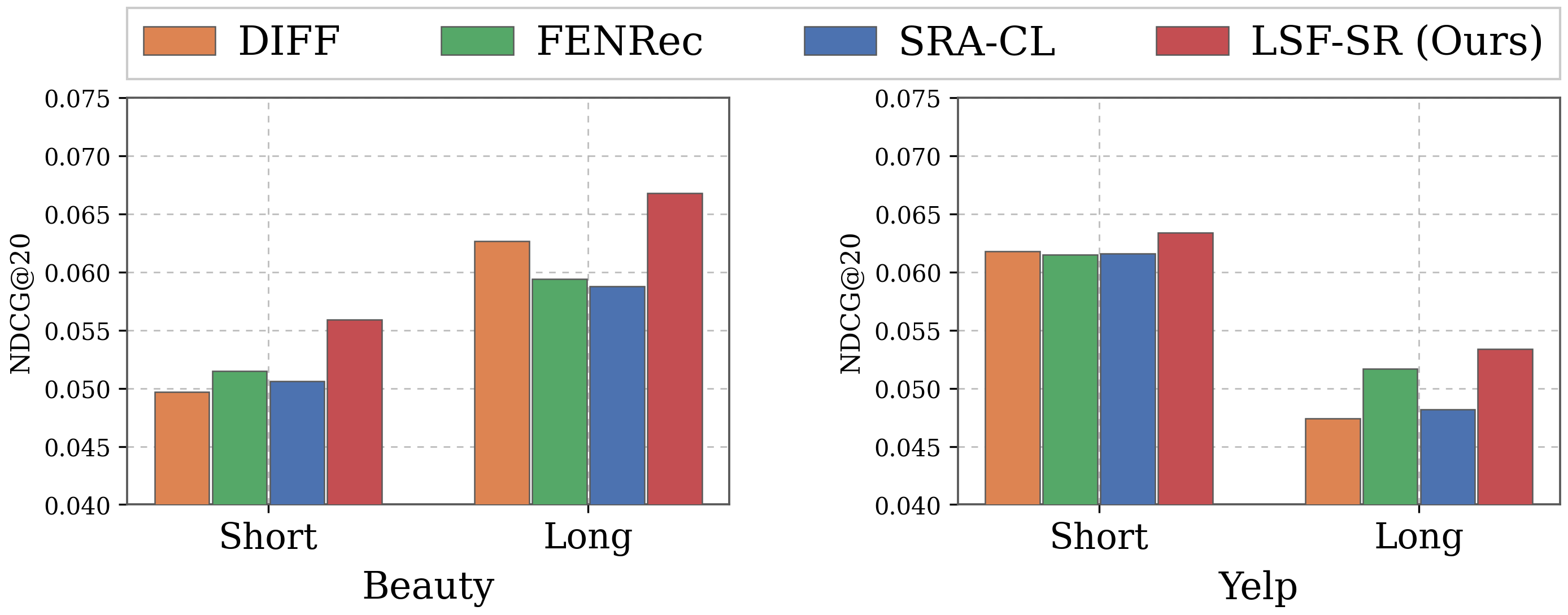}
    \Description{Grouped bar charts displaying the NDCG@20 performance of DIFF, FENRec, SRA-CL, and LSF-SR on Short and Long user sequences across Beauty and Yelp datasets. LSF-SR consistently shows the highest bars in all groups.}
    \caption{Performance comparison (NDCG@20) across user groups with varying sequence lengths. Short sequences are defined as those with a length less than or equal to 5.}
    \label{fig:case_study_seq}
\end{figure}

\begin{figure}[t]
    \centering
    \includegraphics[width=\columnwidth]{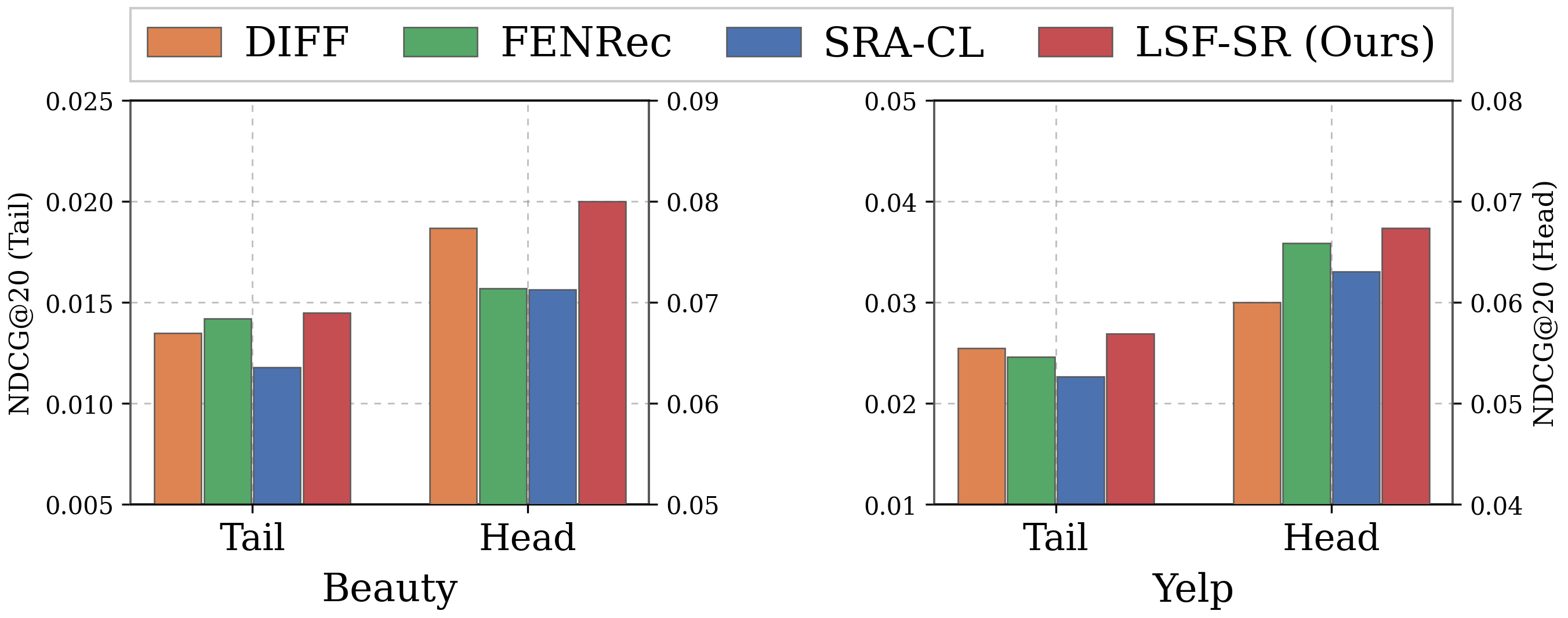}
    \Description{Grouped bar charts showing the NDCG@20 performance of various models on Tail and Head items for Beauty and Yelp datasets. Dual Y-axes separate the numerical scales for Tail and Head groups. LSF-SR consistently outperforms the baselines.}
    \caption{Performance comparison (NDCG@20) across target items with varying popularity levels. Tail items are defined as those appearing at most 9 times. Note that dual y-axes are employed to accommodate the differences in scale between the tail and head groups.}
    \label{fig:case_study_item}
    \vspace{-4pt}
\end{figure}

To assess the robustness of our framework under severe data sparsity and cold-start scenarios, we conducted an in-depth case study using the representative Beauty and Yelp datasets.
We divided the test cases using the first quartile (Q1) as the statistical threshold: user sequences were categorized into two groups—short sequences (length $\le 5$) and long sequences.
Similarly, we classified ground-truth items into two groups—tail items (frequency $\le 9$) and head items.
Because accurate ranking is most important when interaction signals are scarce, we report performance comparisons using NDCG@20, as shown in Figures~\ref{fig:case_study_seq} and~\ref{fig:case_study_item}.

In short interaction sequences, LSF-SR consistently outperforms all representative baselines.
An interesting pattern in the Yelp dataset is that all models perform better on short sequences than on long ones.
This contrasts with e-commerce domains, such as Beauty, where long sequences can reflect stable user preferences.
In Yelp interactions, however, users often explore local businesses and dining options episodically.
As a result, short sequences tend to capture cohesive, immediate intents, such as searching for similar restaurants in a specific neighborhood over a weekend.
In contrast, long sequences span longer periods and often include significant location shifts and intent changes, making next-item prediction more challenging and noisy.

Despite the focused nature of these short-term intents, the scarcity of historical interactions remains a major obstacle. 
Behavior-centric methods struggle here because they lack sufficient co-occurrence patterns. 
Additionally, existing LLM-augmented methods that rely on independent retrieval strategies often suffer from semantic misalignment. 
In contrast, our proposed probabilistic fusion approach uses rich textual semantics as robust conditions, enabling the framework to accurately identify immediate user intents. 
This allows us to maintain superior ranking quality, even with extremely limited behavioral data.

Furthermore, LSF-SR shows significant advantages in recommending tail items, which often suffer from the most severe lack of collaborative signals.
The results show that LSF-SR achieves remarkable improvements in NDCG@20, increasing by 22.88\% and 18.50\% over the strongest LLM-augmented baseline (SRA-CL) for tail items in the Beauty and Yelp datasets, respectively.
Even with advanced techniques---including time-dependent soft labeling and coarse-to-fine intent contrastive learning in FENRec, or frequency filtering and side-information integration in DIFF---these baselines still underperform our semantic-conditioned fusion approach.
Tail-item embeddings trained only on discrete identifiers are often undertrained because interactions are scarce.
By generating item descriptions with an LLM and producing enhanced embeddings through a flow-augmented conditional variational module, LSF-SR can place tail items by semantic relatedness rather than popularity, improving ranking in cold scenarios. 

\section{Conclusion}
\label{sec:conclusion}
In this paper, we present \textbf{LSF-SR}, a novel probabilistic fusion framework for sequential recommendation that recasts heterogeneous alignment as a stochastic inference problem. 
Unlike prior approaches that rely on deterministic projections or late fusion, \textbf{LSF-SR} employs a Conditional Variational Autoencoder in which LLM-derived semantic embeddings condition the transformation of collaborative item representations into a shared latent space. 
To capture the non-Gaussian geometry of the joint collaborative--semantic distribution, we augment the posterior with stacked planar or radial normalizing flows, enabling fine-grained distributional modeling beyond what a diagonal Gaussian posterior can express. 
A dynamic KL annealing strategy further stabilizes training and prevents the latent fusion path from collapsing to an unregularized deterministic mapping. 
Extensive evaluations across five benchmark datasets demonstrate that \textbf{LSF-SR} consistently outperforms state-of-the-art baselines from both behavior-centric and LLM-augmented paradigms, achieving up to 12.98\% in Recall@20 and 14.13\% in NDCG@20. 
On tail items, the framework achieves relative NDCG@20 improvements of 22.88\% and 18.50\% over the strongest LLM-augmented baseline on Beauty and Yelp, respectively.
Case studies under extreme data sparsity and cold-start scenarios further establish the practical robustness of \textbf{LSF-SR}. 
In future work, we plan to explore adaptive flow architectures that dynamically calibrate posterior expressiveness based on domain-specific data characteristics, as our ablation results suggest that datasets with sparser collaborative signals derive comparatively less benefit from deep flow stacks. 
Additionally, the intent drift patterns observed in the Yelp case study motivate the development of temporally aware semantic conditioning mechanisms that can track evolving user preferences over extended interaction horizons.

\section*{Acknowledgments}
This research was supported in part by the National Science and Technology Council Taiwan under grant nos. 115-2223-E-A49-001 and 115-2221-E-005-094-MY3.

\section*{GenAI Usage Disclosure}
During the preparation of this work, the authors used generative artificial intelligence (AI) tools solely for language editing and improving the manuscript's grammatical clarity.
The authors did not use generative AI tools for experimental design, code implementation, data analysis, or developing the core algorithms and methodology.
The authors have thoroughly reviewed and edited the final text and take full responsibility for the academic integrity and content of this paper.


\bibliographystyle{ACM-Reference-Format}
\bibliography{sample-base}


\end{document}